\documentclass[a4paper,11pt]{article}
\pdfoutput=1
\usepackage{jheppub}
\usepackage{mathtools}
\usepackage{physics}
\usepackage{multirow}
\usepackage[normalem]{ulem}
\usepackage{comment}

\def\micro{{\tt micrOMEGAs}}
\def\sigmaCC{{\tt SigmaCC}}

\newcommand{\gs}{g_\star}
\newcommand{\gss}{g_{\star s}}
\newcommand{\Trh}{T_\text{rh}}

\title{\boldmath Testing frozen-in pNGB dark matter\\with a long-lived dark Higgs}
\author[a]{Nicolás Bernal,}
\emailAdd{nicolas.bernal@nyu.edu}
\author[b,c]{Giovanna Cottin,}
\emailAdd{gfcottin@uc.cl}
\author[c,b]{Bastián Díaz Sáez,}
\emailAdd{bastian.diaz@uc.cl}
\author[b,c]{Manuel López}
\emailAdd{manuel.lopez.f@uc.cl}

\affiliation[a]{New York University Abu Dhabi\\
PO Box 129188, Saadiyat Island, Abu Dhabi, United Arab Emirates}
\affiliation[b]{Instituto de Física, Pontificia Universidad Católica de Chile\\
Avenida Vicuña Mackenna 4860, Santiago, Chile}
\affiliation[c]{Millennium Institute for Subatomic Physics at the High-Energy Frontier (SAPHIR)\\
Fernández Concha 700, Santiago, Chile}

\abstract{We consider a simple Higgs portal dark matter (DM) model, where the Standard Model is extended with a complex singlet scalar. The imaginary part of the scalar becomes a massive and stable pseudo-Nambu-Goldstone boson, serving as the DM candidate, while the real part gives rise to a second (dark) Higgs boson. We focus on the freeze-in production of the DM, paying particular attention to low-reheating temperature scenarios, where the dark Higgs can be a long-lived particle (LLP). We also explore the phenomenology of this dark Higgs at the LHC and the Future Circular Collider in hadron-hadron mode, discussing its discovery prospects in regions of parameter space consistent with current DM constraints. Our results emphasize the impact of the cosmic reheating dynamics on the DM freeze-in production, and their critical role in interpreting collider signatures. Furthermore, our findings suggest that LLP searches may provide insights into the fundamental dynamics of reheating.}

\begin{document} 
\begin{flushright}
\end{flushright}

\maketitle

\section{Introduction}
Since the discovery of the Higgs boson in 2012, the door to new physics at the electroweak scale has remained open. However, the absence of clear signals at colliders and direct detection experiments suggests that if this new physics exists, it might involve a sector that couples only feebly to the Standard Model (SM). Dark matter (DM) stands as a prime candidate for this new physics, but traditional WIMP scenarios are increasingly in tension with recent experimental bounds from direct detection~\cite{Arcadi:2024ukq, Cirelli:2024ssz, EscuderoAbenza:2025cfj}. This motivates the exploration of alternative models where the DM particle resides in a potentially rich hidden sector, weakly connected to the SM, which could still be accessible at current or future experiments (see, for instance, Refs.~\cite{Chu:2011be, Bharucha:2022lty, Batell:2022dpx, Harris:2022vnx}).

An attractive framework for this purpose is freeze-in DM production~\cite{McDonald:2001vt, Choi:2005vq, Kusenko:2006rh, Petraki:2007gq, Hall:2009bx, Elahi:2014fsa, Bernal:2017kxu}, particularly in scenarios with low reheating temperature $\Trh$~\cite{Dolgov:1989us, Traschen:1990sw, Kofman:1994rk, Kofman:1997yn, Allahverdi:2020bys, Batell:2024dsi, Barman:2025lvk}. In this regime, the interaction rates controlling the coupling between the hidden sector and the SM can be significantly larger, making it possible for such models to be probed in laboratory settings. In fact, a growing body of work has recently focused on the interplay between low $\Trh$ freeze-in, direct detection, and/or prompt collider physics~\cite{Giudice:2000ex, Fornengo:2002db, Pallis:2004yy, Gelmini:2006pw, Drees:2006vh, Yaguna:2011ei, Roszkowski:2014lga, Drees:2017iod, Bernal:2018ins, Bernal:2018kcw, Cosme:2020mck, Arias:2021rer, Bernal:2022wck, Bhattiprolu:2022sdd, Haque:2023yra, Chowdhury:2023jft, Boddy:2024vgt, Ghosh:2023tyz, Cosme:2023xpa, Silva-Malpartida:2023yks, Barman:2024nhr, Barman:2024lxy, Arias:2023wyg, Arcadi:2024wwg, Bernal:2024yhu, Silva-Malpartida:2024emu, Arcadi:2024obp, Barman:2024tjt, Bernal:2024ndy, Lee:2024wes, Belanger:2024yoj, Bernal:2025fdr, Khan:2025keb, Mondal:2025awq, Bernal:2025osg, Bernal:2025szh, Bernal:2025fcl}. 

Although the phenomenology of long-lived particles~\cite{Curtin:2018mvb, Alimena:2019zri}, usually studied in the context of standard freeze-in, that is, high $\Trh$, has received a lot of attention in recent years (see Ref.~\cite{Westhoff:2023xho} for a short summary), low $\Trh$ freeze-in scenarios are also worth exploring, since higher couplings can not only keep the long-liveness of neutral or charged BSM mediators~\cite{Belanger:2018sti, Calibbi:2018fqf, Junius:2019dci, No:2019gvl, Calibbi:2021fld, Becker:2023tvd, Arias:2025nub}, but eventually these scenarios can be complemented with other types of searches, for example direct detection. In this way, it is exciting to explore more deeply the phenomenology of low $\Trh$ freeze-in DM scenarios, in particular the one appearing from a long-lived dark Higgs boson.

In this work, we consider a minimal extension of the SM with a complex singlet scalar field, which acquires a vacuum expectation value (VEV)~\cite{Gross:2017dan} (see also Ref.~\cite{Coito:2021fgo} and references therein). This setup naturally leads to two new physical degrees of freedom: a stable pseudo-Nambu Goldstone boson (pNGB) as a DM candidate and a second Higgs boson (from now on dark Higgs). We study the freeze-in production of this pNGB in the early universe for low $\Trh$, focusing on the impact of the singlet-Higgs mixing angle, which not only governs the freeze-in efficiency but also directly influences the lifetime of the dark Higgs. This connection allows us to link the cosmological history of DM production with long-lived particle (LLP) searches for the first time in this specific setup.

The paper is organized as follows. In Section~\ref{sec:model}, we present our model setup and parameter spectrum choices. In Section~\ref{sec:fi}, we discuss the mechanisms that contribute to the freeze-in production of DM. Section~\ref{sec:constraints} reviews the experimental constraints on the model. In Section~\ref{sec:pheno}, we analyze the LLP phenomenology of our dark Higgs. In Section~\ref{sec:results} we present our main results. Finally, we summarize our conclusions in Section~\ref{sec:conclusions}.

\section{Model}\label{sec:model}
We add to the SM a complex singlet scalar $S$ charged under a new global $U(1)$ symmetry explicitly broken by a soft term~\cite{Gross:2017dan}. At the renormalizable level, the scalar potential is
\begin{equation}\label{pot1}
    V(H , S) = -\frac{\mu_H^2}{2}|H|^2 - \frac{\mu_S^2}{2}| S |^2 + \frac{\lambda_H}{2}| H |^4 + \frac{\lambda_S}{2}| S |^4 + \lambda_{HS}| H|^2 |S|^2 + V_{\text{soft}}\,,
\end{equation}
with $\lambda_H > 0$, $\lambda_S > 0$, $H$ the Higgs doublet, and the soft-breaking term given by
\begin{equation}\label{soft}
    V_{\text{soft}} = -\frac{m_\chi^2}{4}\left(S^2 + {S^*}^2\right),
\end{equation}
where $m_\chi > 0$. This soft-breaking term lifts the would-be Goldstone boson associated with the spontaneous breaking of the global $U(1)$ symmetry, giving it a nonzero mass. Choosing $\mu_S^2 < 0$ and rewriting the complex field as $S = (s' + i\, \chi)/\sqrt{2}$, the vacuum is located at ${s'}^2 + \chi^2 = v_s^2 > 0$. Without loss of generality, we take $\langle s' \rangle = v_s$ and $\langle \chi \rangle = 0$. Although the global $U(1)$ is explicitly broken by the soft term, the full scalar potential remains invariant under the discrete transformation $S \to S^*$, which in terms of the real fields corresponds to
\begin{equation}
    s' \to s' \qquad \text{and} \qquad \chi \to -\chi\,.
\end{equation}
This defines a remnant $\mathbb{Z}_2$ symmetry under which only the pseudoscalar component $\chi$ is odd. Since the vacuum preserves this symmetry, it remains unbroken after spontaneous symmetry breaking. Furthermore, the presence of this $\mathbb{Z}_2$ ensures the stability of $\chi$, making it a viable candidate for DM.

It is convenient to rewrite the fields around their VEVs and set $s' = v_s + s$ so that we have
\begin{equation}
    H = \frac{1}{\sqrt{2}} \begin{pmatrix} 0 \\ v_h + h \end{pmatrix} \qquad \text{and} \qquad S = \frac{1}{\sqrt{2}} (v_s + s + i\chi)\,.
\end{equation}
We demand the potential to be stationary with respect to fluctuations of the real scalar fields $h$ and $s$ at the vacuum
\begin{equation}
    \left. \frac{\partial V}{\partial h} \right|_{h = s = 0} = 0 \qquad \text{and} \qquad \left. \frac{\partial V}{\partial s} \right|_{h = s = 0} = 0\,,
\end{equation}
which implies that
\begin{align}
    \mu_H^2 &= \lambda_H v_h^2 + \lambda_{HS}v_s^2 \,, \\
    \mu_S^2 &= \lambda_{HS} v_h^2 + \lambda_{S}v_s^2 - m_\chi^2 \,.
\end{align}
The mass matrix of the states $(h,s)$ is given by
\begin{equation} \label{mass}
\mathcal{M}^2 =
    \begin{pmatrix}
        \lambda_h\, v_h^2 & \lambda_{hs}\, v_h\, v_s \\
        \lambda_{hs}\, v_h\, v_s & \lambda_s\, v_s^2 
    \end{pmatrix} ,
\end{equation}
whereas the mass of $\chi$ is simply given by $m_\chi$. The mass matrix in Eq.~\eqref{mass} can be diagonalized using a set of orthogonal matrices as diag$(m_1^2,\, m_2^2)= \mathcal{O}\, \mathcal{M}^2\, \mathcal{O}^T$, with  
\begin{equation}
    \mathcal{O} =
    \begin{pmatrix}
        \cos\theta & -\sin\theta \\
        \sin\theta & \cos\theta
    \end{pmatrix} .
\end{equation}
The mass eigenstates $h_1$ and $h_2$ relate to the interaction eigenstates $h$ and $s$ such as $h = \cos\theta\, h_1 + \sin\theta\, h_2$ and $s = -\sin\theta\, h_1 + \cos\theta\, h_2$, and the masses of the new CP-even scalars are
\begin{equation}\label{masses}
    m_{h_1,\, h_2}^2 = \frac{1}{2} \left(\lambda_S\, v_s^2 + \lambda_H\, v_h^2 \mp \frac{\lambda_S\, v_s^2 - \lambda_H\, v_h^2}{\cos 2\theta}\right).
\end{equation}
A useful relation among the parameters is given by 
\begin{equation} 
  \tan 2 \theta = \frac{2\lambda_{HS}\, v_h\, v_s}{\lambda_S\, v_s^2 - \lambda_H\, v_h^2}
\end{equation}
which, in the limit of small mixing angles $\theta \ll 1$, reduces to
\begin{equation}
    \lambda_{HS} \simeq \frac{m_{h_2}^2 - m_{h_1}^2}{v_h\, v_s}\, \theta\,.
\end{equation}
From Eq.~\eqref{masses}, we can also express the quartic couplings as
\begin{align} \label{quartic3}
    \lambda_H &= \frac{1}{2\, v_h^2} \left[m_{h_1}^2 + m_{h_2}^2 + \cos 2\theta \left(m_{h_1}^2 - m_{h_2}^2\right)\right], \\
    \lambda_S &= \frac{1}{2\, v_s^2} \left[m_{h_1}^2 + m_{h_2}^2 - \cos 2\theta \left(m_{h_1}^2 - m_{h_2}^2\right)\right].
\end{align}

Given the previous expressions, and after identifying $h_1$ as the SM-like Higgs boson (with $v_h \simeq 246$~GeV and $m_{h_1} \simeq 125$~GeV), we choose {$\left\{ m_\chi,\, m_{h_2},\, v_s,\, \theta \right\}$} as the free parameters. Table~\ref{tab:scan_ranges} shows the ranges for the parameters considered. These parameters are chosen so that we can focus on the dark matter and collider phenomenology of a long-lived dark Higgs boson.
\begin{table}[h]
    \centering
    \begin{tabular}{c c}
    \hline
    Parameter & Scan range \\
    \hline
    $m_\chi$ & $[3~{\rm GeV},\, m_W)$ \\
    $m_{h_2}$ & $[3~{\rm GeV},\, m_W)$ \\
    $\sin\theta$ & $[10^{-6},\, 1]$ \\
    $v_s$ & $100~{\rm GeV}$ (fixed) \\
    \hline
    \end{tabular}
    \caption{Parameter ranges used in our numerical scan. The vacuum expectation value $v_s$ was fixed for simplicity, while the remaining parameters were varied within the indicated intervals.}
    \label{tab:scan_ranges}
\end{table}

Our parameter scan focuses on the most relevant region for obtaining a long-lived dark Higgs, which could be probed at the LHC or future colliders such as the FCC-hh. This requirement naturally restricts both the scalar and dark matter masses to lie below the electroweak scale, motivating the adopted ranges for $m_\chi$ and $m_{h_2}$. The lower limit of 3~GeV is chosen to avoid potential complications from hadronic resonances in the GeV range~\cite{Winkler:2018qyg}, which could affect the dark matter production mechanisms. As we later use some approximations in the calculation of the relic abundance, we keep the upper bounds of the masses below $m_W$ for simplicity—namely, to avoid the opening of additional annihilation channels. Similarly, the chosen interval for the mixing angle $\sin\theta$ covers values for which the decay length of the dark Higgs becomes large enough (from $\mathcal{O}(\text{mm})$ to several meters) to be potentially detectable with long-lived particle search strategies at colliders. The choice of $v_s = 100$~GeV is not crucial for collider phenomenology and was fixed close to the electroweak scale for definiteness. It certainly impacts dark matter production, but varying it would not qualitatively change our main results. Throughout this work, we also impose the condition $m_{h_2} < 2m_\chi$ to focus on the region where the dark matter analysis simplifies (requiring only one Boltzmann equation), and where the decays of $h_{2}$ into fermions can produce electrically charged particles displaced from the primary proton–proton collision vertex. The complementary case $m_{h_2} > 2m_\chi$, in which invisible decays become relevant, is left for future investigation.

\section{Dark matter production}\label{sec:fi}
The evolution of the DM number density $n_\chi$ can be followed by the use of the Boltzmann equation
\begin{equation} \label{boltzeq1}
    \frac{dn_\chi}{dt} + 3\, H\, n_\chi = \mathcal{C}\,,
\end{equation}
where the Hubble expansion rate $H$, in a universe dominated by SM radiation, is\footnote{We note that we do not need to fix an inflationary scale. For the values of $\Trh$ considered in our work ($\Trh \sim \mathcal{O}$(GeV)), we only require for the initial Hubble scale $H_I \gtrsim 10^{-16}$~GeV.}
\begin{equation}
    H(T) = \frac{\pi}{3}\, \sqrt{\frac{\gs(T)}{10}}\, \frac{T^2}{M_P}\,,
\end{equation}
with $M_P \simeq 2.4 \times 10^{18}$~GeV the reduced Planck mass, $T$ the temperature of the SM bath, and $\gs(T)$ the number of relativistic degrees of freedom contributing to the SM energy density. Furthermore, the DM production rate density $\mathcal{C}$ is given by
\begin{equation} \label{eq:collision}
    \mathcal{C} = \mathcal{C}_{h_1 \to \chi\chi} + \mathcal{C}_{h_2 \to \chi\chi} + \sum_{f\in \text{fermions}} \mathcal{C}_{f\bar{f} \to \chi\chi} + \sum_{X\in \text{bosons}} \mathcal{C}_{X\bar{X} \to \chi\chi}\,,
\end{equation}
and contains the following tree-level contributions: direct decay of the Higgs bosons $h_1$ and $h_2$, annihilations of SM fermions, SM gauge bosons, and Higgs bosons through the $s$-channel exchange of $h_1$ and $h_2$, and the contact interaction. We note that since we are considering $m_{h_2} < 2\, m_\chi$, the decay of $h_2$ into a pair of DM particles is kinematically closed. Additionally, even if annihilations of SM gauge bosons and Higgs bosons are allowed, their contributions are subdominant because of a Boltzmann suppression in the parameter space of our interest, in which their masses are much higher than the maximal temperature reached by the SM thermal bath. Therefore, we will focus on the production of DM from the decay of $h_1$ and the annihilation of light SM fermions; see Fig.~\ref{fig:diagrams}.\footnote{For completeness, in Appendix~\ref{app:thermalization} we verify that in this scenario DM never reaches thermal equilibrium, so that it is a FIMP. Additionally, we note that DM self-interactions are present, but are not strong enough to bring the dark sector into chemical equilibrium with itself~\cite{Bernal:2015xba, Bernal:2020gzm}.}
\begin{figure}[t!]
  \centering
  \includegraphics[width=0.8\textwidth]{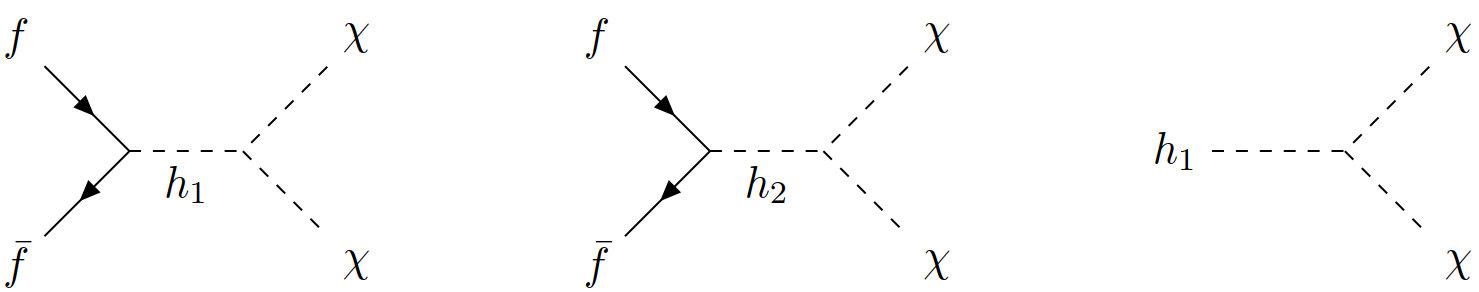}
  \caption{Main processes contributing to the DM production in the early Universe.}
  \label{fig:diagrams}
\end{figure}

The corresponding DM production rates can be calculated taking into account that~\cite{Gondolo:1990dk}
\begin{equation}
    \mathcal{C}_{X\to \chi\chi} \simeq \int \left(\prod_{i \in X} d\Pi_i\right) d\Pi_{\chi_1}\, d\Pi_{\chi_2}\, (2\pi)^4\, \delta^{(4)}(p_X - p_{\chi_1} - p_{\chi_2}) \left(\prod_{i \in X} f_i\right) |\mathcal{M}(X \to \chi \chi)|^2,
\end{equation}
where $X$ corresponds to the initial state particles, $d\Pi_i$ is the Lorentz-invariant phase space, and $f_i$ the phase-space distribution functions. We can work out the DM production rates in each case of Eq.~\eqref{eq:collision} giving rise to\footnote{We assume a Maxwell–Boltzmann distribution for all particles in the thermal bath, which is a good approximation for freeze-in at low reheating temperatures, where dark matter is produced non-relativistically.}
\begin{align} \label{c1}
  \mathcal{C}_{h_1 \rightarrow \chi\chi}(T) &= 2\, \frac{K_1(m_{h_1}/T)}{K_2(m_{h_1}/T)}\, \Gamma_{h_1 \rightarrow \chi\chi}\, n_{h_1}^\text{eq} \simeq 2\, \Gamma_{h_1 \rightarrow \chi\chi}\, n_{h_1}^\text{eq}\,,\\
  \mathcal{C}_{f\bar{f} \rightarrow \chi\chi}(T) &= \ev{\sigma v}_{f\bar{f} \rightarrow \chi\chi}\, \left({n_f^\text{eq}}\right)^2 = \ev{\sigma v}_{\chi\chi \rightarrow f\bar{f}}\, \left({n_{\chi}^\text{eq}}\right)^2\,,\label{c2}
\end{align}
with $\Gamma_{h_1 \rightarrow \chi\chi}$ the decay width of the 125 GeV Higgs boson into two DM particles (see Appendix~\ref{A}), and the equilibrium number densities given by 
\begin{equation}\label{nie}
    n_X^\text{eq}(T) =  g_{X}\, \frac{m_{X}^2}{2\pi^2}\, T\, K_2\left(\frac{m_{X}}{T}\right),
\end{equation}
where $X$ represents a specific particle, $g_{X}$ the internal degrees of freedom of $X$, and $K_i$ are the modified Bessel function of the second kind. In the second equality in the collision term in Eq.~\eqref{c2}, we have used the detailed balance principle (see, e.g. Ref.~\cite{Bringmann:2021sth}). For practical reasons, it is advantageous to compute $\mathcal{C}_{f\bar{f} \rightarrow \chi\chi}$ from the inverse reaction to obtain an analytical expression~\cite{Arcadi:2024obp}, since we are focused on the parameter space where $m_\chi \gtrsim m_f$, for all SM fermions but the top quark. In addition, since we always have $\Trh \ll m_\chi$, we can safely expand $(\sigma v)_{\chi\chi \rightarrow f\bar{f}}$ in powers of $v$ and, retaining the leading term in the expansion, $\ev{\sigma v}_{\chi\chi \rightarrow f\bar{f}} \approx (\sigma v)_{\chi\chi \rightarrow f\bar{f}}|_{s=4m_\chi^2}$ (see Appendix~\ref{A}). Notice that the contribution from the top quark is absolutely subleading for $\mathcal{C}_{f\bar{f} \rightarrow \chi\chi}$ in the parameter space that we are focused on, since at low $T$ its contribution is highly Boltzmann suppressed.

Plugging the corresponding expressions into the RHS of Eqs.~\eqref{c1} and~\eqref{c2}, we obtain the following rates
\begin{align}
    \mathcal{C}_{h_1\to \chi\chi}(T) &\simeq \frac{\sin^2\theta}{32\, \sqrt{2}\, \pi^\frac52}\, \frac{m_{h_1}^\frac92\, T^\frac32}{v_s^2}\, \left[1 - \left(\frac{2\, m_\chi}{m_{h_1}}\right)^2\right]^\frac12 e^{-\frac{m_{h_1}}{T}}, \label{coll1}\\
    \mathcal{C}_{f\bar{f} \to \chi\chi}(T) &\simeq \frac{N_c\, \sin^2\theta}{2\pi^3} \left[\frac{m_{h_1}^2 - m_{h_2}^2}{(m_{h_1}^2 - 4m_\chi^2)\, (m_{h_2}^2 - 4m_\chi^2)}\right]^2 \frac{m_f^2\, m_\chi^7\, T^3}{v_h^2\, v_s^2} \left[1 - \left(\frac{m_f}{m_\chi}\right)^2\right]^\frac32 e^{-\frac{2m_\chi}{T}},\label{coll2}
\end{align}
where the exponential factors reflect the Boltzmann suppression of the present scenario with low-temperature reheating, where $m_{h_1} > \Trh$ and $m_\chi > \Trh$, and $N_c =1$ for leptons or $N_c =3$ for quarks is the color factor.

For convenience, let us define the dimensionless DM yield $Y \equiv n_\chi/s$, where
\begin{equation}
    s(T) = \frac{2\, \pi^2}{45}\, \gss(T)\, T^3
\end{equation}
is the SM entropy density and $\gss(T)$ is the number of relativistic degrees of freedom that contribute to the SM entropy. Now, Eq.~\eqref{boltzeq1} can be analytically solved as
\begin{equation} \label{yield}
  Y(T) \simeq \int_T^{\Trh} dT'\, \frac{\mathcal{C}(T')}{H(T')\, s(T')\, T'}\,,
\end{equation}
where the reheating temperature $\Trh$ corresponds to the highest temperature reached by the SM thermal bath in the instantaneous reheating approximation. We note that contributions to the DM density generated in the cosmic inflationary era, that is, before $T = \Trh$, were neglected. Before solving the integral to obtain each yield in Eq.~\eqref{yield}, we note that the main contribution to the integral comes from temperatures close to $\Trh$; therefore, we ignore the temperature dependence of $\gs$ and $\gss$, and we evaluate them at $T = \Trh$.

If DM is mainly produced by Higgs decays, Eq.~\eqref{yield} can be analytically solved as
\begin{equation} \label{eq:Y0dec}
    Y_0^d \simeq \frac{135\, \sqrt{5}\, \sin^2\theta}{128\, \gss\, \sqrt{\gs}\, \pi^\frac{11}{2}}\, \frac{m_{h_1}^\frac72\, M_P}{v_s^2\, \Trh^\frac52} \left[1 - \left(\frac{2\, m_\chi}{m_{h_1}}\right)^2\right]^\frac12 e^{-\frac{m_{h_1}}{\Trh}},
\end{equation} 
where we have considered $\Trh \ll m_{h_1}$. Alternatively, if production is dominated by SM-particle scatterings, Eq.~\eqref{yield} gives
\begin{align} \label{eq:Y0ann}
    Y_0^s &\simeq \frac{405\, \sqrt{5}\, \sin^2\theta}{24\, \sqrt{2\, \gs}\, \gss\, \pi^7} \frac{(m_{h_1}^2 - m_{h_2}^2)^2}{\left[(m_{h_1}^2 - 4m_\chi^2)^2 + m_{h_1}^2\Gamma_{h_1}^2\right]\, \left[(m_{h_2}^2 - 4m_\chi^2)^2 + m_{h_2}^2\Gamma_{h_2}^2\right]} \frac{m_\chi^5\, M_P}{v_h^2\, v_s^2\, \Trh} \nonumber\\
    &\qquad\qquad \times\sum_f  N_c\, m_f^2\, (2m_\chi + \Trh) \left[1 - \left(\frac{m_f}{m_\chi}\right)^2\right]^\frac32 e^{-\frac{2m_\chi}{\Trh}},
\end{align}
where we have included the total decay width for the propagators of $h_1$ and $h_2$. In general, the two contributions of Eqs.~\eqref{eq:Y0dec} and~\eqref{eq:Y0ann} should be taken into account: $Y_0 = Y_0^d + Y_0^s$.

To match the entire observed DM relic density at present, it is required that
\begin{equation} \label{relic-density-no}
    m\, Y_0 = \frac{\Omega h^2\, \rho_c}{s_0\, h^2} \simeq 4.3 \times 10^{-10}~\text{GeV},
\end{equation}
where $Y_0$ is the asymptotic value of the DM yield at low temperatures, $s_0 \simeq 2.69 \times 10^3$~cm$^{-3}$ is the present entropy density~\cite{ParticleDataGroup:2024cfk}, $\rho_c \simeq 1.05 \times 10^{-5}~h^2$~GeV/cm$^3$ is the critical energy density of the Universe and $\Omega h^2 \simeq 0.12$ is the observed abundance of DM relics~\cite{Planck:2018vyg}.

\begin{figure}[t!]
  \centering
  \includegraphics[width=0.45\textwidth]{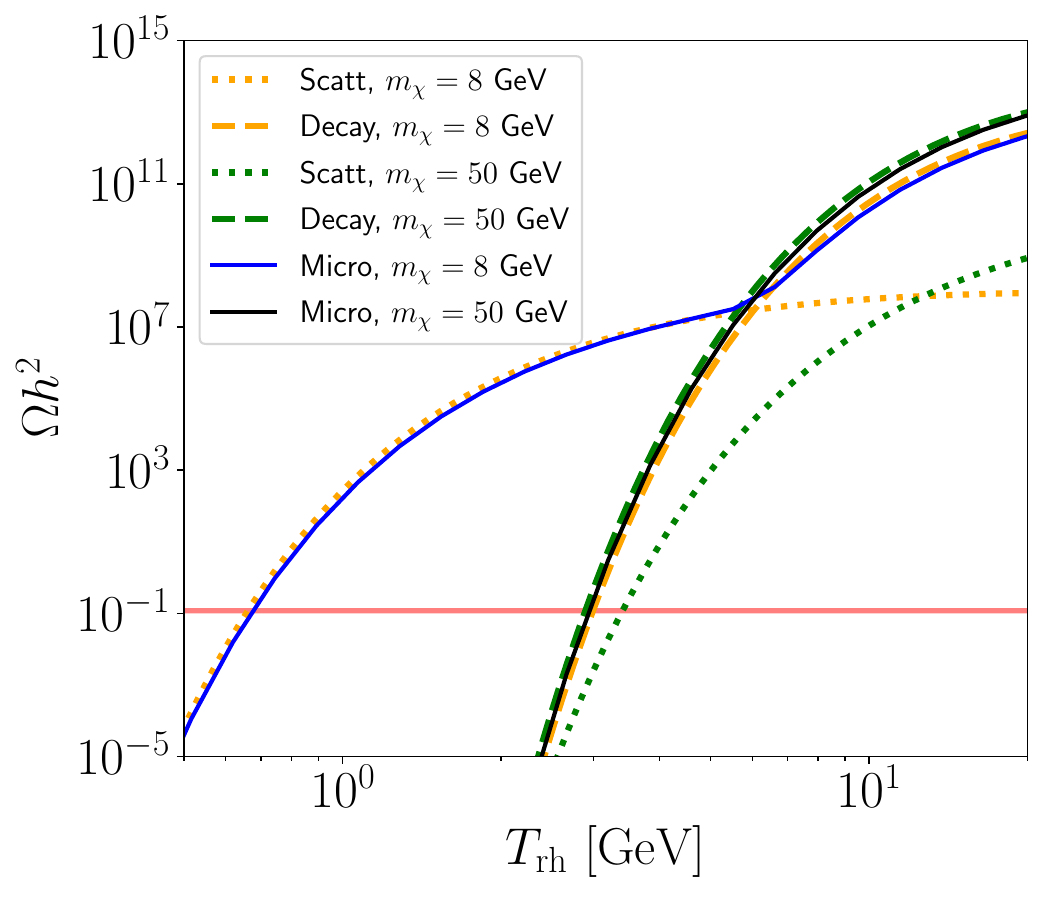}
  \includegraphics[width=0.45\textwidth]{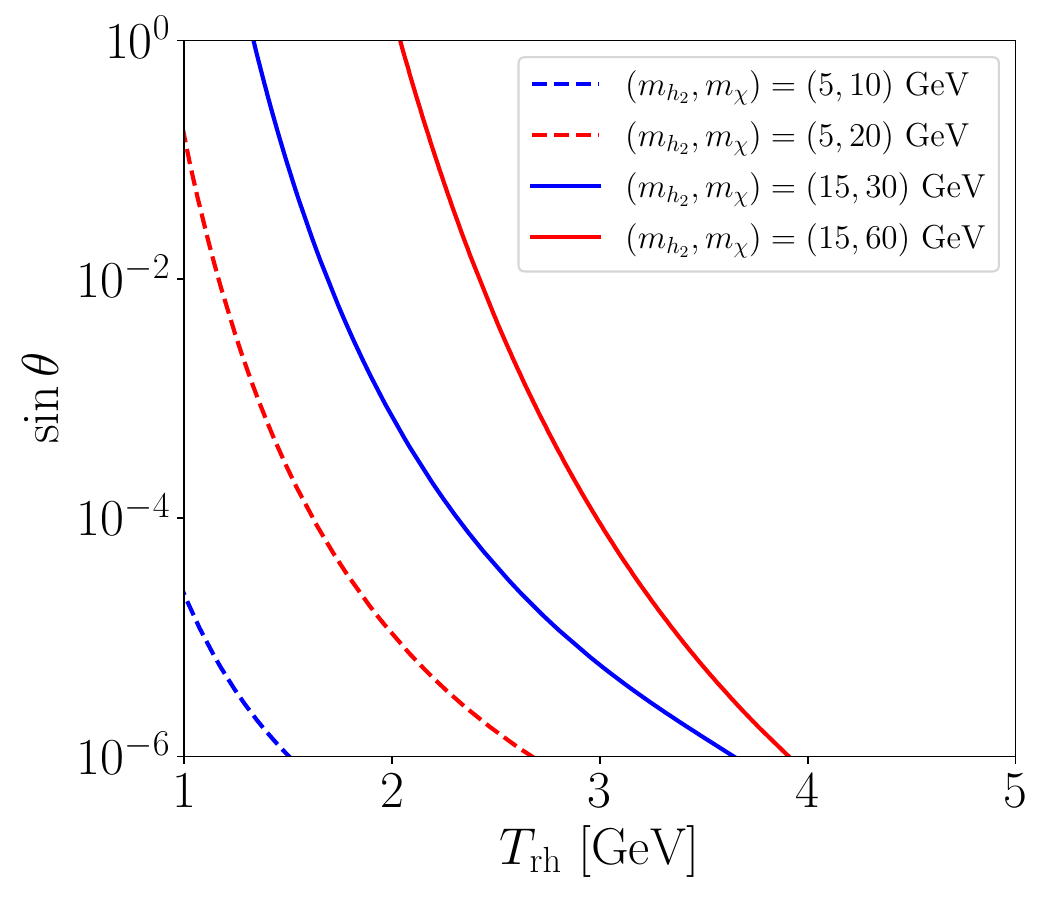}
  \caption{Left panel: Relic abundance as a function of the reheating temperature $\Trh$, considering $m_{h_2} = $ 10~GeV, $m_\chi = $ 8~GeV and $m_\chi = $ 50~GeV, $\sin\theta = 10^{-4}$ and $v_s = $ 100~GeV. We have separated the freeze-in contributions from Higgs decay (dashed lines) and scattering (dotted lines) contributions obtained analytically in this section for the two DM masses shown in the legend. The blue and black lines are the corresponding \micro\ result, which give the full relic abundance. Right panel: Required values of $\sin\theta$ to fit the observed relic abundance, as a function of $\Trh$, for different combination of the pair $(m_{h_2}, m_\chi)$, and $v_s = 100$~GeV.}
  \label{plot_relic}
\end{figure}
The left panel of Fig.~\ref{plot_relic} shows the DM relic abundance, $\Omega h^2$, as a function of the reheating temperature $\Trh$, for $m_{h_2} = 10$~GeV, $\sin\theta = 10^{-4}$, $v_s = 100$~GeV, and two DM benchmark values $m_\chi = 8$~GeV and $m_\chi = 50$~GeV. For each DM mass, we display the partial contributions to the relic abundance from annihilations (dotted lines) and decays (dashed lines), while the solid blue and solid black lines correspond to the total relic density obtained with \micro~6.0.4~\cite{Belanger:2018ccd}. The first point to emphasize is the excellent agreement between the analytical results of Eqs.~\eqref{eq:Y0dec} and~\eqref{eq:Y0ann}, together with the \micro\ computation (differences at the percent level). We also observe that the correct relic abundance can be achieved for both $m_\chi$ values, although for different reheating temperatures and dominant freeze-in mechanisms—either scatterings or decays.\footnote{We have verified with \micro\ that for $\Trh \gg m_\chi$, the relic abundance becomes independent of $\Trh$, as expected in the IR freeze-in regime.} Naturally, for $m_\chi > m_{h_1}/2$, decays are kinematically forbidden and only scatterings contribute to freeze-in, typically requiring larger values of $\theta$ to reproduce the observed relic abundance (as shown explicitly later).

Furthermore, the values of $\sin\theta$ required to reproduce the observed total DM abundance exhibit a strong dependence on $\Trh$ for fixed model parameters. This behavior is shown in the right panel of Fig.~\ref{plot_relic} for different combinations of $m_\chi$ and $m_{h_2}$, with $v_s = 100$~GeV. As expected, the suppression of the DM yield at low $\Trh$ must be compensated by a significant increase in the mixing angle $\sin\theta$.

\begin{figure}[t!]
  \centering
  \includegraphics[width=0.5\textwidth]{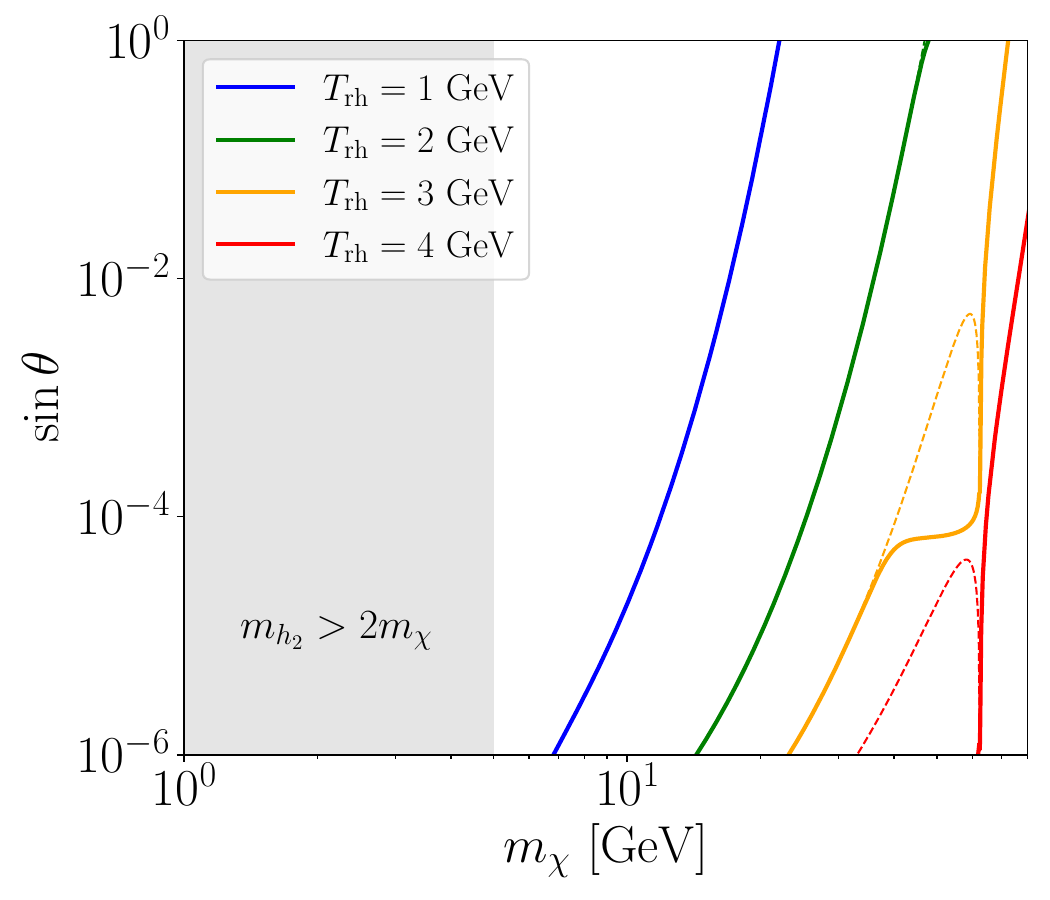}
  \caption{Parameter space fitting the observed DM relic abundance, assuming $m_{h_2} = 10$~GeV and $v_s = 100$~GeV, for different reheating temperatures (thick solid lines). The thin dashed lines take into account only the contribution from scatterings.}
  \label{plot_relic2}
\end{figure}
Figure~\ref{plot_relic2} shows the parameter space in the $[m_\chi,, \sin\theta]$ plane that reproduces the observed DM relic abundance, assuming $m_{h_2} = 10$~GeV and $v_s = 100$~GeV, for different reheating temperatures. For large reheating temperatures, DM is produced mainly through Higgs decays of $h_1$, and very small mixing angles are required to ensure FIMP production. This channel becomes kinematically closed near $m_\chi = m_{h_1}/2$. Lower reheating temperatures require larger values of the mixing angle. Furthermore, as shown in Fig.~\ref{plot_relic}, annihilations become dominant for smaller values of $\Trh$. We also include the pure scattering contribution in the absence of decays (dashed lines) when it reproduces the correct relic abundance, in order to illustrate the deviations that arise once decays are included. Finally, the region on the left, where $m_\chi < m_{h_2}/2$, is not considered since in that case DM could be produced by decays of $h_2$.

Having cross-checked our analytical expressions for the relic abundance with \micro, we conclude this section by commenting on the role of $v_s$ in the parameter space, which has been fixed to 100~GeV throughout our analysis. In principle, $v_s$ could take higher values. However, if the goal is to reproduce the correct relic abundance while remaining within our target region of the parameter space, $[m_\chi, m_{h_2}, \theta]$, the reheating temperature $\Trh$ must increase accordingly to compensate for the suppression induced by a larger $v_s$. This behavior arises because the DM yield scales inversely with $v_s$, making DM production less efficient as $v_s$ increases. In such cases, a higher $\Trh$ would require the calculation of more annihilation channels, as noted in Section~\ref{sec:fi}. Conversely, smaller values of $v_s$ would allow for a lower $\Trh$, although in that regime unitarity bounds become relevant. In both limits, we have verified with \texttt{micrOMEGAs} that the correct relic abundance can still be obtained, although we did not explore these regions of parameter space in detail.

\section{Constraints} \label{sec:constraints}
In this section, we discuss and present relevant bounds on this Higgs portal model, coming from DM direct detection and collider searches.

\subsection*{Direct detection}
Pseudo-Nambu-Goldstone boson DM presents a natural suppression in the spin-independent rate with nuclei, since the scattering amplitude is proportional to the small transfer momentum, thereby not imposing any relevant constraint~\cite{Gross:2017dan, Abe:2021vat}.\footnote{An exception to this can be realized by making $m_{h_2} \ll 1$ GeV, such that the dark Higgs mass is comparable to the transfer momentum $|q|$ as shown in Ref.~\cite{Abe:2021vat}, even for values of $\theta$ in the ballpark of values that we are considering in this work (e.g. $\theta \sim 10^{-4}$). However, here we consider $m_{h_2} \gtrsim 1$~GeV.}

\subsection*{Indirect Detection Constraints}
Annihilations of DM $\chi\chi \to h_2 h_2$ into real or virtual mediators that subsequently decay into SM particles can give rise to observable indirect detection (ID) signatures. The strongest bounds come from \textsc{Fermi}-LAT observations of gamma rays from dwarf spheroidal galaxies~\cite{Fermi-LAT:2015att}. In the case where $m_\chi > m_{h_2}$, the effective annihilation cross section is given by~\cite{Abe:2021vat}
\begin{equation}\label{eq:sigmaveff}
    \langle \sigma v\rangle_{\rm eff} = \langle \sigma v\rangle_{\chi\chi \to h_2 h_2}\, \times [{\rm BR}(h_2\to f\bar f)]^2,
\end{equation}
and can be compared with the upper bounds from  Ref.~\cite{Elor:2015bho}, where ${\rm BR}(h_2\to f\bar f)$ is the branching ratio of $h_2$ to a pair of fermions $f \bar f$. In our parameter space, $h_2$ decays mainly into $\tau\bar{\tau}$ and $b\bar{b}$ with a branching ratio of approximately 0.2 and 0.7, respectively. In the case of $m_\chi < m_{h_2}$, we obtain the corresponding four-body $\langle \sigma v\rangle_{\rm eff}$ directly from \micro\ making use of the function \sigmaCC. The resulting constraints are presented in Figure~\ref{plot_ctau2}, at the end of the following section.

\subsection*{Meson decays}
Dark Higgses with masses above the GeV scale can be produced at the radiative level in $B$ meson decays. Here, we use the bounds in the plane $[m_{h_2},\, \sin\theta]$~\cite{Ferber:2023iso}, based on the LHCb data from $B \rightarrow K^*\, h_2 \rightarrow K^*\, \mu^+\mu^-$~\cite{LHCb:2015nkv, LHCb:2016awg}.

\subsection*{Prompt Collider bounds}
Dark Higgs bosons can be probed through precise measurements of the SM-like Higgs boson properties and direct searches at colliders or beam dump experiments. In the region of interest for our study, $1 \text{ GeV} \lesssim m_{h_2} \lesssim m_{h_1}/2$, the most stringent limits arise from the former, specifically from measurements of the Higgs signal strength (HSS), which quantify deviations in the combination of production and decay channels of a 125~GeV Higgs boson from SM predictions~\cite{Bhattiprolu:2022ycf, Ferber:2023iso}. For our case, the HSS is given by
\begin{equation}\label{eq:mu}
    \mu \equiv \frac{\sigma}{\sigma_{\text{SM}}}\frac{\text{Br}_j}{\text{Br}_{\text{SM},j}}\,,
\end{equation}
where $\sigma$  is the production cross section of $h_1$, related to the production cross section of the 125 GeV Higgs in the SM (i.e. only a Higgs doublet in the scalar sector of the SM) by $\sigma = \cos^2\theta\, \sigma_{\text{SM}}$. Similarly, $\text{Br}_j$ corresponds to the branching ratio of the $h_1\rightarrow j^\text{th}$ SM final states, while $\text{Br}_{\text{SM},j}$ is the same branching fraction for the SM Higgs. In our case, the HSS is given by
\begin{equation}
    \mu = \cos^2\theta\, \frac{\Gamma_{h_1}^\text{SM} \cos^2\theta}{\Gamma_{h_1}^\text{SM} \cos^2\theta + \Gamma_{h_1 \rightarrow \text{dark}}}\,,
\end{equation}
with $\Gamma_h^\text{SM} \simeq 4.07$~MeV~\cite{LHCHiggsCrossSectionWorkingGroup:2016ypw}, and $\Gamma_{h_1 \rightarrow \text{dark}}$ considers all possible $h_1$ decays into particles of the dark sector. Taking into account the average of the latest bounds on $\mu$ from ATLAS~\cite{ATLAS:2022vkf} and CMS~\cite{CMS:2022dwd}, the resulting signal strength is given by~\cite{Ferber:2023iso}
\begin{equation}
    \mu = 1.03 \pm 0.06\,.
\end{equation}
Finally, the decay $h_1 \to \chi \chi$ contributes to the invisible decay of the Higgs. ATLAS and CMS have limits on $\text{Br}(h \to \text{inv})$ corresponding to $< 10.7\%$ and $< 15\%$, respectively, at 95\% CL~\cite{ATLAS:2023tkt, CMS:2023sdw}. However, this bound is less restrictive than the one coming from the HSS.

\section{Long-lived particles: collider phenomenology}\label{sec:pheno}
In this section, we study in detail the phenomenology at colliders of the long-lived dark Higgs in this model. We first discuss our model implementation and then perform numerical Monte Carlo simulations. We analyze the LHC limitations to test the model and then estimate prospects at the Future Circular Collider (FCC) in the hadron-hadron mode FCC-hh.

To study the phenomenology of the long-lived Higgs, we first implement the model into {\texttt{FeynRules}}~\cite{Christensen:2008py, Alloul:2013bka}.\footnote{We validate our UFO implementation by recovering results from the LHC Higgs Cross Section Working Group~\cite{Dittmaier:1318996} for different possible values of the mass of the SM Higgs boson.} We focus on gluon-gluon fusion (ggF) as the main production channel $pp\rightarrow h_{2}$.\footnote{In Ref.~\cite{Huitu:2018gbc} it is argued that vector boson fusion (VBF) presents lower backgrounds than ggF, and they estimate a significance factor of VBF over ggF larger by a factor of $\sim 1.6$. Furthermore, they find that the VBF is efficient only when the DM pair is produced by on-shell $h_1$ and/or $h_2$ decays, then they focused on $m_{h_2} > 2\, m_\chi$.} The effective coupling of $h_{1,2}$ to gluons is given by~\cite{Nakamura:2017irk, Cline:2019okt}
\begin{equation}\label{lag_eff}
    \mathcal{L}_\text{eff} =  \frac{\alpha_s\, \cos\theta}{12 \pi\, v_h}\, h_1\, G_{\mu\nu}^a G^{a\mu\nu} +  \frac{\alpha_s\, \sin\theta}{12 \pi\, v_h}\, h_2\, G_{\mu\nu}^a G^{a\mu\nu}.
\end{equation}
As can be inferred from Eq.~\eqref{lag_eff}, the ggF production cross section of $h_2$ is suppressed by the mixing $\sin^2\theta$.

\begin{figure}[t!]
  \centering
  \includegraphics[width=0.5\textwidth]{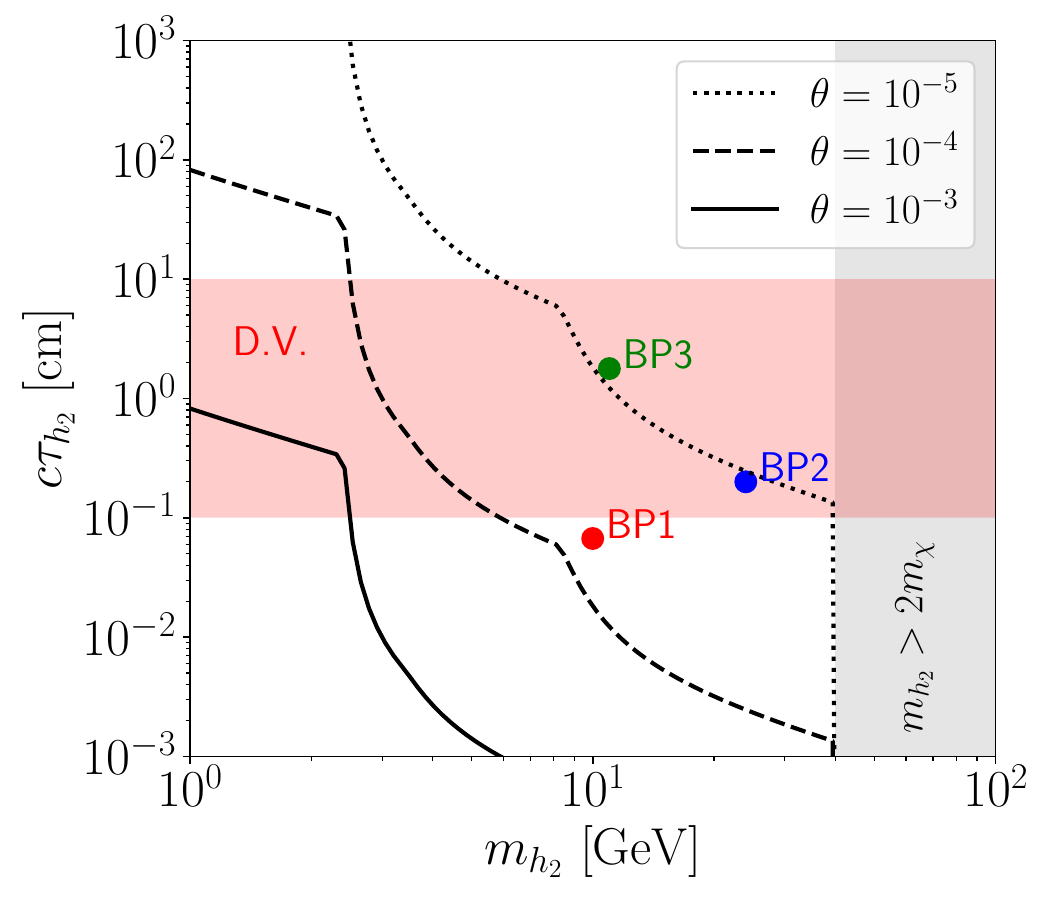}
  \caption{Proper decay length of $h_2$ as a function of $m_{h_2}$, assuming $v_s = 10^2$ GeV and $m_\chi = 20$~GeV. The horizontal pink band (`D.V.')  corresponds to a displacement between 0.1~cm and 10~cm. The red, blue and green points correspond to the benchmark points indicated in Table~\ref{tab:LHC_BM}.}
  \label{plot_ctau1}
\end{figure}
The mixing angle $\theta$ suppresses the production of $h_2$ and its decay width, enhancing its proper decay distance $c\, \tau_{h_{2}}$, as can be seen in Fig.~\ref{plot_ctau1} as a function of $m_{h_2}$. The pink horizontal band (denoted as `D.V.') corresponds to a displacement between 0.1~cm and 10~cm, a representative acceptance of displaced decays inside inner trackers of the main LHC experiments. The relevant parameters for the phenomenology of long-lived dark Higgs are $c\, \tau_{h_{2}}$ and $m_{h_{2}}$. However, we assume $v_s = 10^2$~GeV and $m_\chi = 20$~GeV. We notice that $c\, \tau_{h_{2}}$ is independent of $m_\chi$ if $m_{h_2} < 2\, m_\chi$. Finally, concerning the decay products of the dark Higgs, if its decay channel into DM particles is kinematically closed ($m_{h_2} < 2\, m_\chi$), it can only decay into SM states without any suppression in its branching ratio (for more details, see Appendix~\ref{A}).

\subsection{LHC prospects}
We focus on LHC production at $\sqrt{s} = 14$~TeV with an integrated luminosity $\mathcal{L} = 3000$~fb$^{-1}$. For $h_{2}$ masses above 2~GeV, we computed the expected number of dark Higgses produced that decay into visible states, $N= \mathcal{L} \times \sigma(pp\rightarrow h_{2})\times \text{Br}(h_{2}\rightarrow \text{vis.})$, where $\text{Br}(h_{2}\rightarrow \text{vis.})\sim 0.99$ labels the $h_{2}$ branching ratio into two leptons or $q\bar{q}$ (that is, visible states that are electrically charged). For this computation, we used \texttt{Madgraph5}~\cite{Alwall:2011uj, Alwall:2014hca} and assume an optimistic perfect efficiency of the detectors. Representative benchmarks are presented in Table~\ref{tab:LHC_BM} and Fig.~\ref{plot_ctau1}. These benchmarks were chosen to yield different values of $c\tau_{h_{2}}$ and $N$, in order to understand how small (or large) the proper lifetime has to be for $h_{2}$ to be potentially detected by the LHC main detector trackers. 
\begin{table}[t!]
  \centering
  \begin{tabular}{|c||c|c||c|c|c|}
  \hline
  Benchmark & $\boldsymbol{m_{h_{2}}}$ [GeV] & $\boldsymbol{\theta}$ & $\boldsymbol{\sigma(pp \rightarrow h_{2})}$ [fb] & $\boldsymbol{c\tau_{h_{2}}}$ [mm] & $\boldsymbol{N}$ \\
  \hline\hline
  \bf{BP1} & 10 & $5\times10^{-5}$ & $1.88 \times 10^{-3}$ & $0.67$ & $5.62$ \\
  \hline
  \bf{BP2} & 24 & $1\times10^{-5}$  & $2.81 \times 10^{-5}$ &$2.0$ & $0.08$ \\
  \hline
  \bf{BP3} & 11 & $1\times10^{-5}$  & $6.87 \times 10^{-5}$& $10.78$ & $0.21$ \\
  \hline
  \end{tabular}
  \caption{Representative benchmarks for the LHC at $\sqrt{s}=14$ TeV and $\mathcal{L}=3000$ fb$^{-1}$. }
  \label{tab:LHC_BM}
\end{table}

We note that, even with a naive estimation of $N$, without considering any efficiency effects or geometrical detector acceptances, the suppression of $\sin^2\theta$ in the production limits the number of expected events at the LHC. The more macroscopic the decay length of $h_{2}$ is (which is, in contrast, favored by the small mixing), the less the number of events $N$ occurs. For instance, BP1 in Table~\ref{tab:LHC_BM} is the benchmark we found with the largest $c\tau_{h_{2}}$ that yields $N$ bigger than 3 events at the LHC.\footnote{Under the zero background assumption, by requiring at least $N=3$ one can derive a 95$\%$ CL exclusion limit.} This largest $c\tau_{h_{2}}$ is less than $1$ mm, and so it is right on the edge of the D.V. region (see Fig.~\ref{plot_ctau1}), which we define between 1~mm and 100~mm. Typical requirements for displaced vertices in ATLAS require a cut on the long-lived particle decay between 4 mm and 300 mm, to allow the decay to lay in the bulk of the ATLAS inner tracker~\cite{ATLAS:2017tny}. Even if the boost factor increases the (laboratory) decay length above $\sim1$ mm (and so it could lie inside the red region), this benchmark had the largest $c\tau_{h_{2}}$ we found within our grid, with an $N$ that is not high enough, so we do not consider our relevant region of parameter space promising to be tested at the LHC. Furthermore, cuts and detector simulation will reduce the number of expected events. Following the same arguments, BP2 and BP3 in Table~\ref{tab:LHC_BM} have a larger $c\tau_{h_{2}}$, but a low number of expected events, such as when increasing $m_{h_{2}}$ (or decreasing $\theta$), we lose sensitivity due to a smaller cross section.

All in all, it is very challenging to test the present scenario at the high-luminosity LHC with displaced signatures. In the next section, the possibilities offered by the FCC are studied.

\subsection{FCC prospects}
Given the limited possibilities offered by the LHC, we now study the potential offered by a future hadron collider operating at a higher center of mass energy of $\sqrt{s}=100$~TeV to test this model. For the next generation of colliders at the high-energy frontier, two hadron collider projects are under consideration:\footnote{For a review on current proposed machines operating at the high-precision and energy frontiers, see Ref.~\cite{Maltoni:2022bqs}.} the Future Circular Collider (FCC)~\cite{FCC:2018vvp} and the Super Proton Proton Collider (SPPC)~\cite{Tang:2022fzs}. We focus on FCC in its hadron-hadron mode (FCC-hh) as it is planned to continue with the scientific program of CERN's LHC, as recommended by the European Strategy for Particle Physics Update (ESPPU)~\cite{EuropeanStrategyforParticlePhysicsPreparatoryGroup:2013fia, EuropeanStrategyGroup:2020pow}.\footnote{Details on several FCC-hh feasibility studies can be found in the technical report in Ref.~\cite{FCC:2018vvp}. See also Ref.~\cite{ESPPG:2019qin} for details on the dedicated physics community inputs for the 2020 EPPSU.} Even though no major technical obstacles are foreseen for FCC, the timescales are still not yet certain. Two stages for FCC are planned. First, an electron positron collider, FCC-ee, and 10 years later, a $\sim100$ km circumference proton-proton collider, FCC-hh. FCC-hh would start physics operations in $\sim 2060$~\cite{FCC:2018vvp} (provided FCC-ee is implemented first), reaching an expected integrated luminosity of $\mathcal{L}=20$ ab$^{-1}$. 

\begin{figure}[t!]
  \centering
  \includegraphics[width=0.8\textwidth]{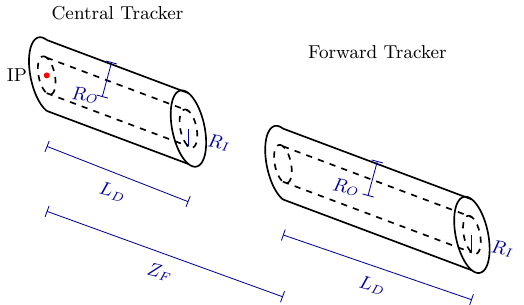} 
  \caption{Schematic overview of the FCC-hh reference detector: geometry and distances.}
  \label{fig:FCChhRef}
\end{figure}

The FCC-hh reference detector setup~\cite{FCC:2018vvp} was proposed based on its robustness and technical feasibility, it is supposed to be comparable to the dimensions of ATLAS or CMS, and includes a forward tracker and a central tracker detector, with both trackers having different geometrical acceptances. This detector would be sensitive to catch displaced decays between $\mathcal{O}(10)$~mm to $\mathcal{O}$(10)~m.\footnote{Note that also several LLP detectors have been proposed at FCC-hh, such as far detectors as DELIGHT and FOREHUNT~\cite{Bhattacherjee:2023plj} or transverse detectors such as the FACET-like and MATHUSLA-like proposals in Ref.~\cite{Boyarsky:2022epg}, which would be suitable for catching lighter (and longer-lived) LLPs produced from the decays of mesons. For other recent long-lived detector proposals at FCC, see also Ref.~\cite{Bhattacherjee:2025dlu}.} In Fig.~\ref{fig:FCChhRef} we show a schematic view of the detector geometry with the definition of the distances that we use for each tracker. In Table~\ref{tab:trackersGeo}, we specify the numerical values for these distances. 

We compute the probability for a decay of a long-lived $h_{2}$ in the central or forward trackers. The probability of decay of an $h_{2}$ within the fiducial volume (f.v.) of a detector (if its flight path traverses the detector) is given by~\cite{deVries:2015mfw, Cottin:2021lzz}
\begin{equation} \label{eq:decayprobability}
    P[ h^{i}_{2} \text{ decay in f.v.} ] = e^{-L^{i}_1/\lambda^i_z} \left(1 - e^{-L^{i}_2/\lambda^i_z}\right),
\end{equation}
where $\lambda^i_z$ corresponds to the mean decay length of the $i$ long-lived $h_{2}$ in the lab frame, which we calculate as~\cite{deVries:2015mfw}
\begin{align}
    \lambda^i_z & \equiv \beta^{i}_{z}\, \gamma\,  c\, \tau_{h_{2}}\,,\\
    \beta^{i}_{z} & = p^{i}_{z}/E^{i}\,,\\
    \gamma^{i} & = E^{i}/m_{h_{2}}\,,
\end{align}
with $c\tau_{h_{2}}=c\hbar/\Gamma(h_{2})_{\text{tot.}}$ (being $\Gamma(h_{2})_{\text{tot.}}$ the total decay width of $h_{2}$), $\beta^{i}_{z}$ the $z$-component of the relativistic velocity of the $i^\text{th}$ $h_{2}$ and $\gamma^{i}$ its corresponding Lorentz boost factor. In Eq.~\eqref{eq:decayprobability}, $L^{i}_1$ corresponds to the distance from the IP to the point where the long-lived $h_{2}$ reaches the detector, $L^{i}_2$ the distance $h_{2}$ traveled through the internal space of the detector. 

\begin{table}[t!]
    \centering
    \begin{tabular}{|c|cccc|}
        \hline
        \multirow{2}{*}{FCC-hh detector} & \multicolumn{4}{c|}{Fiducial decay volume dimensions} \\
         & $\boldsymbol{R_I}$ [mm] & $\boldsymbol{R_O}$ [m] & $\boldsymbol{L_D}$ [m] & $\boldsymbol{Z_F}$ [m]\\
        \hline
        {\bf Central tracker} & 20 & 1.7 & 5 & - \\
        {\bf Forward tracker} & 20 & 1.6 & 6 & 10 \\
        \hline
    \end{tabular}
    \caption{Relevant distances for the central and forward FCC-hh trackers~\cite{FCC:2018vvp}. $R_{I}$ corresponds to the beam pipe inner radius. $R_O$ is the radius of the corresponding tracker cavity. $L_{D}$ is the length of the tracker. $Z_F$ corresponds to the distance between the IP and the beginning of the forward tracker.}
    \label{tab:trackersGeo}
\end{table}
The relevant geometric parameters that describe the central and forward trackers are provided in Table~\ref{tab:trackersGeo}. For the central tracker, we define the distances as
\begin{align}
  L^{i}_{1} &= \min(L_{D},|R_{I}/\tan{\theta^{i}}|)\,,\\
  L^{i}_{2} &= \min(L_{D},|R_{O}/\tan{\theta^{i}}|) - L^{i}_{1}\,,
  \label{eq:decayprobabilityDistancesCentral}
\end{align}
where $\theta^{i}$ is the polar angle of the $i^\text{th}$ $h_{2}$, calculated as $\theta^{i}=\arctan(p^{i}_{T}/p^{i}_z)$. For the forward tracker, we define the distances as
\begin{align}
    L^{i}_{1} &=
    \begin{dcases}
        |R_{I}/\tan{\theta^{i}}| & \text{if } L_{D}+Z_F >|R_{I}/\tan{\theta^{i}}|\,, \\
        Z_F & \text{if } L_{D}+Z_F <|R_{I}/\tan{\theta^{i}}| \,,
    \end{dcases}\\
    L^{i}_{2} &=
    \begin{dcases}
        \min(L_{D} + Z_F ,|R_{O}/\tan{\theta^{i}}|) & \text{if } L_{D}+Z_F >|R_{I}/\tan{\theta^{i}}| - L^{i}_{1} \,,\\
        Z_F & \text{if } L_{D}+Z_F <|R_{I}/\tan{\theta^{i}}| - L^{i}_{1}\,.
    \end{dcases}
  \label{eq:decayprobabilityDistancesForward}
\end{align}
We also require an acceptance cut for the pseudorapidity $|\eta|<2.5$ for the central tracker and $|\eta|<3$ for the forward tracker~\cite{FCC:2018vvp}. If an $h_{2}$ falls outside the geometrical acceptance of the tracker, $P[h^{i}_2 \text{decay in f.v.}] = 0$.

In order to compute the distances, kinematic variables, and the decay probability, we perform a custom made \texttt{Python} code that reads the four-momenta information from the (parton-level) LHE event files~\cite{Boos:2001cv} generated with \texttt{Madgraph5}~\cite{Alwall:2011uj, Alwall:2014hca}. We compute the relevant cross-sections at $\sqrt{s}=100$~TeV in \texttt{Madgraph5}. In addition to generating $pp\rightarrow h_{2}$, we also add the process $pp\rightarrow h_{2} + X$, with $X$ being any particle emitted in the event to have kinematic information in the transverse plane at the parton level. We can then calculate the number of expected events in each FCC-hh tracker as~\cite{Cottin:2021lzz}
\begin{equation}
    N =  \sigma \times \mathcal{L} \times  \langle P[ h_2 \text{ decay in f.v.} ] \rangle  \times \text{Br}(h_2 \to \text{vis.})\,,
\end{equation}
with
\begin{equation}
    \langle P[ h_2 \text{ decay in f.v.} ] \rangle  \equiv \frac{1}{k} \sum_{i=1}^{k} P[ h^{i}_2 \text{ decay in f.v.}]
\end{equation}
being the average decay probability of all the simulated $h_{2}$ within the fiducial volume of each tracker, $k = 10^5$ is the total number of Monte Carlo simulated events and $\text{Br}(h_{2} \rightarrow \text{vis.})$ corresponds to the branching ratio of $h_{2}$ into visible particles (i.e. electrically charged).

\subsection{Results} \label{sec:results}
\begin{figure}[t!]
  \centering
   \includegraphics[width=0.6\textwidth]{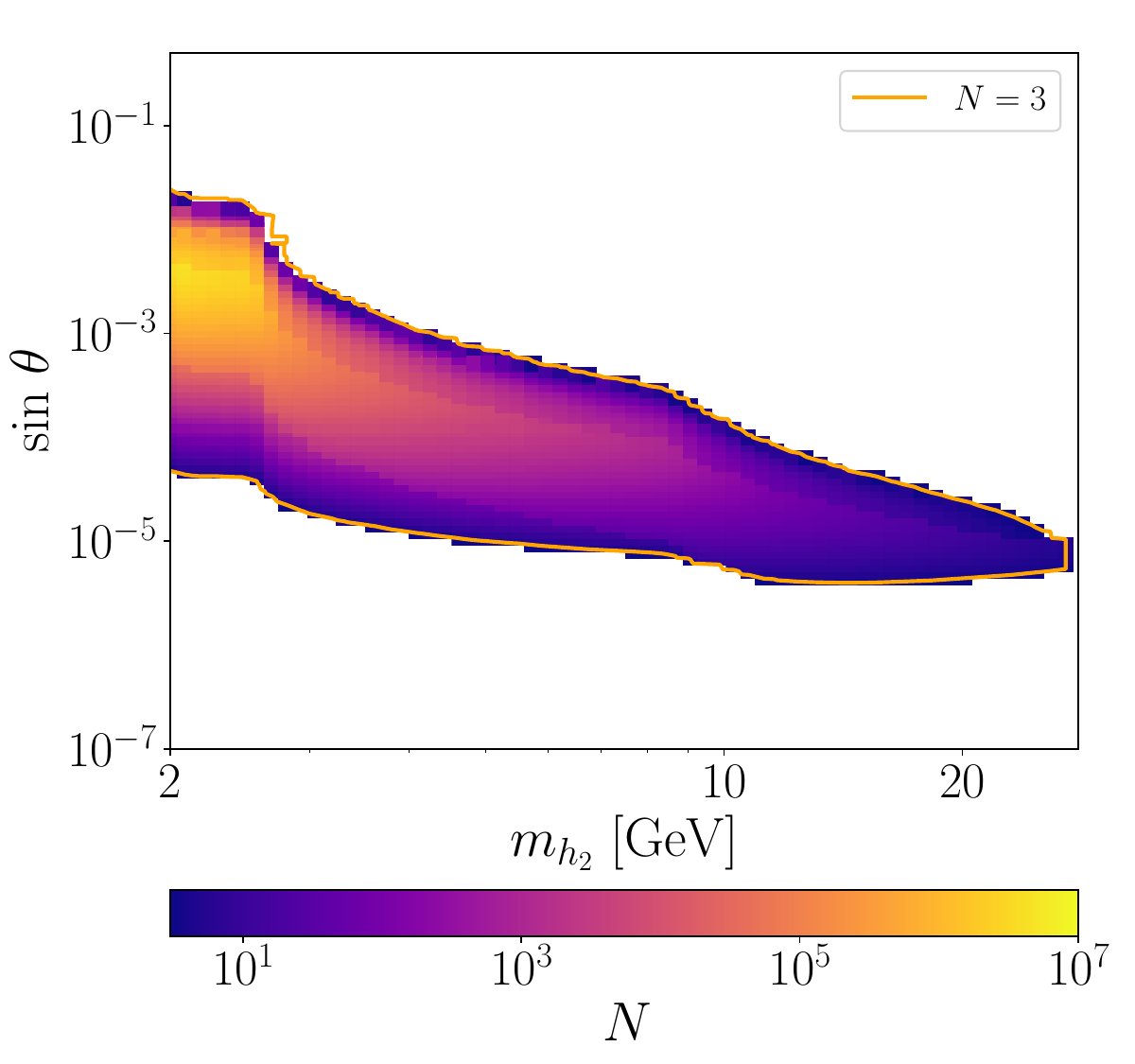} 
    \caption{Number of expected events $N$ at the FCC-hh central and forward trackers combined. The yellow line corresponds to $N=3$ events.}
    \label{fig:FCChhSensitivity}
\end{figure}
Figure~\ref{fig:FCChhSensitivity} shows the number of events expected in the FCC-hh reference tracker (central and forward). Here we consider an integrated luminosity of $\mathcal{L}=20$ ab$^{-1}$, as is expected for the projected lifetime of FCC-hh~\cite{FCC:2018vvp}. By requiring at least 3 signal events, under the zero background assumption, we derive a $95\%$ CL exclusion limit and present it as a contour in the plot. Although the assumption of zero background events is optimistic, ongoing LHC displaced vertex searches report $\sim 0.002$ background events at $\mathcal{L} = 32.8$~fb$^{-1}$~\cite{ATLAS:2017tny}, and less than 1 event at $\mathcal{L} = 139$~fb$^{-1}$~\cite{ATLAS:2023oti}. Furthermore, the nature of these backgrounds is instrumental (i.e. interactions with the detector material), so we cannot estimate them accurately. The FCC-hh reference detector may also not be in its final design stage. Our prospects then serve as a rough estimate where we compare the sensitivity of the FCC-hh reference detector geometry and acceptance. However, one could understand the effect of smaller efficiencies and/or potential non-negligible backgrounds by placing an exclusion contour at higher $N$. We finally note that the shape of the exclusion contour is bounded by $h_{2}$ either decaying too promptly or too far away (i.e. falling outside any of the tracker detector's acceptances). The dips in the shape at around 4~GeV and 10~GeV correspond to the opening of the Higgs decays to charm quarks and tau leptons, and to $b$-quarks, respectively.

\begin{figure}[t!]
  \centering
  \includegraphics[width=1\textwidth]{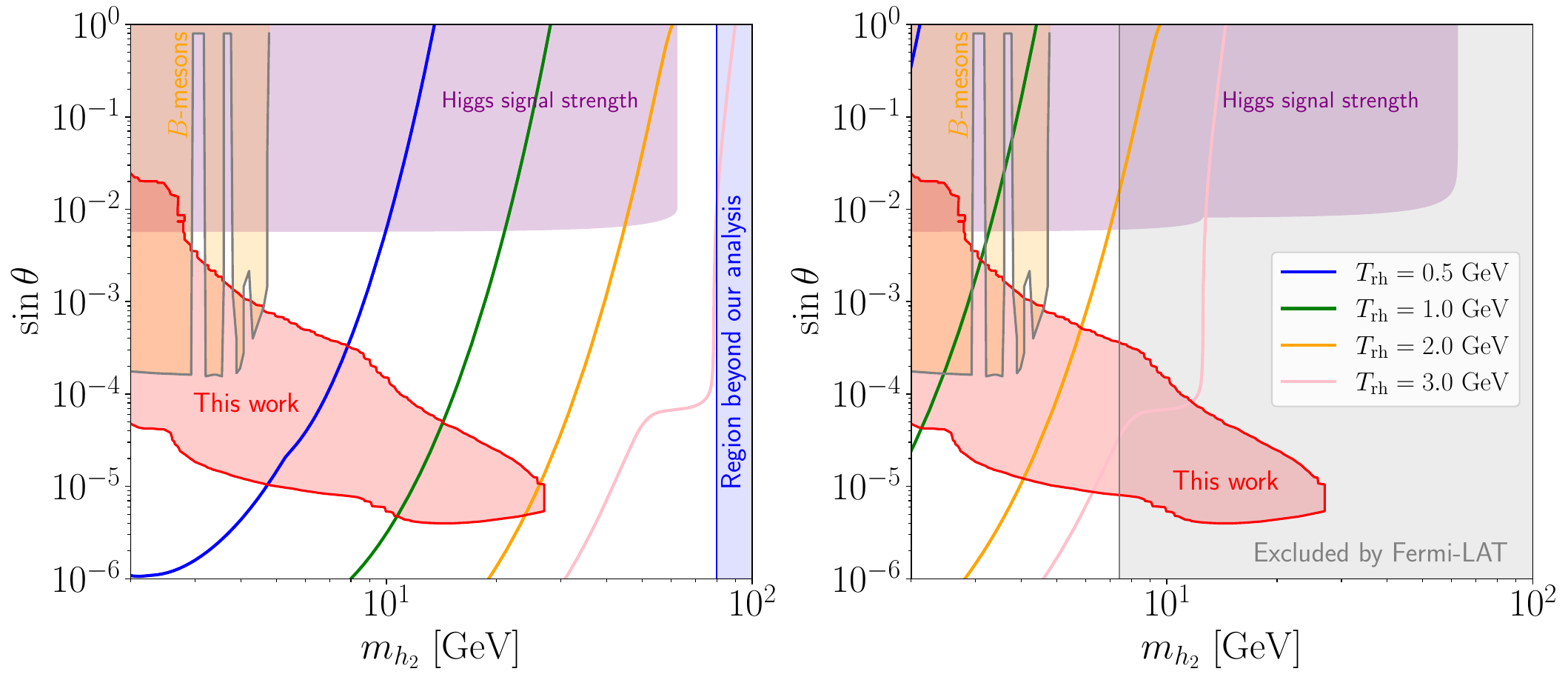} 
  \caption{The color lines show the parameter space that fits the total observed DM relic abundance for different values of $\Trh$, and $m_\chi = 0.8\, m_{h_2}$ (left) and $m_\chi = 5\, m_{h_2}$ (right). The red regions represent the future sensitivity from the forecast of FCC-hh at 100 TeV; cf. Fig.~\ref{fig:FCChhSensitivity}. The purple and orange regions consider the constraints coming from the Higgs signal strength and LHCb from searches of $B$-meson decays, respectively. The gray region on the right panel corresponds to the exclusion expected with indirect detection. The blue region is beyond the analysis carried out in this work, since $m_{h_2} > m_W$. The legend on the right plot applies for both panels.} 
  \label{plot_ctau2}
\end{figure}

In order to confront these 95\% CL contours with the correct relic abundance results based on DM freeze-in at low $\Trh$, in Fig.~\ref{plot_ctau2} we show our final results. The solid lines show different contours of the correct relic abundance for fixed $\Trh$, with the mass of the DM fixed to $m_\chi = 0.8\, m_{h_2}$ (left panel) and $m_\chi = 5\, m_{h_2}$ (right panel). The red regions represent the 95\% CL projected sensitivity obtained in our LLP analysis at the 100~TeV FCC-hh; cf. Fig.~\ref{fig:FCChhSensitivity}. In Fig.~\ref{plot_ctau2} we also included bounds from the Higgs signal strength and $B$-meson decay, and bounds from ID from \textsc{Fermi}-LAT searches (see Section~\ref{sec:constraints}). We note that ID bounds are only relevant as long as $m_\chi > m_{h_2}$ (gray region in the right panel of Fig.~\ref{plot_ctau2}), since it is in this case where $\ev{\sigma v}_\text{eff}$ (see Eq.~\eqref{eq:sigmaveff}) takes sizable values to be contrasted with upper bounds from~\cite{Elor:2015bho}. We have checked with the function \sigmaCC\ of \micro\ 6.2.4 that for $m_\chi < m_{h_2}$ (left panel), the average annihilation cross section is highly suppressed for the $2\rightarrow 4$ process (more than ten orders of magnitude smaller than $\ev{\sigma v}_\text{eff}$ as $m_\chi > m_{h_2}$) due to the appearance of off-shell propagators, thus not imposing any significant constraint. As shown, FCC-hh could test some low reheating scenarios which give the correct relic abundance, complementing already existing constraints.

\section{Conclusions}\label{sec:conclusions}
In this work, we have explored the connection between scalar dark matter (DM) production via freeze-in in low-reheating scenarios and the phenomenology of long-lived particles. Our framework relies on a simple Higgs portal extension of the Standard Model, involving a complex scalar field that acquires a vacuum expectation value, leading to the emergence of a pseudo-Nambu-Goldstone boson and a second, lighter Higgs-like scalar.

Unlike the standard freeze-out scenario, here we have focused on FIMP production during a cosmological era where reheating is instantaneous and occurs at a temperature smaller than the DM mass. Therefore, DM production occurs at temperatures below its mass, significantly reducing the corresponding production rates. This suppression has to be counterbalanced by an increase of the production cross section, which, in turn, requires the singlet-doublet mixing angle to be raised by several orders of magnitude compared to those values required in standard reheating scenarios. Interestingly, this allows the dark Higgs boson to have a very rich LLP phenomenology, as its typical displacement could be within the reach of the LHC and future collider such as the FCC.

While the current capabilities of the LHC are insufficient to probe this parameter space due to the suppressed production cross section of the dark Higgs, we have demonstrated that a 100 TeV future circular collider such as the FCC-hh, operating at higher luminosities, could effectively test the GeV mass range.  Future direct detection DM experiments, such as the DARWIN/XLZD experiment~\cite{Baudis:2024jnk, XLZD:2024nsu}, could overlap the timescales of FCC-hh, as it is intended to start operations around 2030. Nevertheless, we do not expect future direct detection experiments to place relevant bounds, given the pNGB nature of the DM.

In certain regions of parameter space, indirect detection plays an important role in complementing future collider tests of the model. This offers a promising path toward uncovering the nature of DM and its possible connections to hidden, long-lived states at colliders. Our results emphasize the impact of the cosmic reheating dynamics post-inflation on the generation of DM FIMPs and their critical role in interpreting collider signatures. Furthermore, our findings suggest that LLP searches may provide insights into the fundamental dynamics of reheating.

Future studies could extend this framework by considering alternative production and decay modes of $h_{2}$, such as $h_{1}\rightarrow h_{2}h_{2}$, where searches for exotic Higgs decays~\cite{ATLAS:2025xmb} could be reinterpreted for this scenario. In addition, alternative mass hierarchies, such as $m_{h_2} > 2\, m_\chi$, could significantly alter the production channels and the phenomenology of the dark sector. Furthermore, exploring the sub-GeV mass range for dark Higgs could reveal interesting connections to direct detection experiments, where even small couplings might leave observable imprints. Maintaining the focus on low-reheating scenarios also opens up a wide array of possibilities for probing DM across multiple current and upcoming experiments, potentially uncovering previously inaccessible regions of parameter space.

\section*{Acknowledgements}
NB received funding from the grants PID2023-151418NB-I00 funded by MCIU/AEI/10.13039 /501100011033/ FEDER and PID2022-139841NB-I00 of MICIU/AEI/10.13039/501100011033 and FEDER, UE. GC acknowledges support from ANID FONDECYT grant No.~1250135. BDS was funded by ANID FONDECYT postdoctorado 2022 No.~3220566 and ANID FONDECYT Iniciación No.~11251818. We also acknowledge support from ANID – Millennium Science Initiative Program ICN2019\_044. 

\appendix
\section{Cross sections and decay formulas} \label{A}
In the narrow-width approximation, the production of $h_2$ with its subsequent decay into SM particles is given by
\begin{equation}
    \sigma = \sigma(pp\rightarrow h_2) \times \text{Br}(h_2\rightarrow \text{SM})\,,
\end{equation}
where
\begin{equation}
    \sigma(pp\rightarrow h_2) = \sin^2\theta \times \sigma_{pp\rightarrow h_1}^\text{SM}(m_{h_1} = m_{h_2})
\end{equation}
and
\begin{equation}
    \text{Br}(h_2 \rightarrow \text{SM}) = \frac{\sin^2\theta\, \Gamma_{h_2\rightarrow \text{SM}}}{\sin^2\theta\, \Gamma_{h_2} + \Gamma_{h_2\rightarrow \text{dark}}}
\end{equation}
where $\Gamma_{h_2}$ is the full width of $h_2$. As we focus on $m_{h_2} < m_\chi/2$ and $m_{h_2} < m_{h_1}$, the branching fraction of $h_2$ into any set of SM final state does not suffer suppression from the mixing angle, unlike its production cross section.

As the expressions for $h_1$ and $h_2$ are similar for their decays, we present only those for $h_1$, and the cases for the decays of $h_2$ only require changing $m_{h_1} \rightarrow m_{h_2}$ and $\sin\theta(\cos\theta)\rightarrow \cos\theta(\sin\theta)$.  In the case of decay into SM fermions, 
\begin{equation}
    \Gamma_{h_1\rightarrow f\bar{f}} = \frac{G_\mu N_c}{4\sqrt{2}\pi}m_{h_1}m_f^2 \cos^2\theta\left(1 - \frac{4m_f^2}{m_{h_1}^2}\right)^{3/2},
\end{equation}
where $G_\mu \simeq 1.16\times 10^{-5}$ GeV$^{-2}$, and $N_c = 3$ is the color factor in the case of quarks. Alternatively, in the case of a decay into a pair of pNGBs
\begin{equation}
    \Gamma_{h_1\rightarrow \chi\chi} = \frac{m_{h_1}^3\sin^2\theta}{32\pi v_s^2}\sqrt{1 - \frac{4m_\chi^2}{m_{h_1}^2}}\,.
\end{equation}
Another relevant decay in this work is the decay of $h_1$ into a pair of $h_2$. In the limit where $m_{h_2} \ll m_{h_1}$ and $\theta \ll 1$~\cite{Abe:2021vat}
\begin{equation}
    \Gamma_{h_1\rightarrow h_2h_2}  = \frac{m_{h_1}^3\sin^2\theta}{32\pi v_s^2}\sqrt{1 - \frac{4m_{h_2}^2}{m_{h_1}^2}}\,.
 \end{equation}

Finally, the relevant average cross section times relative velocity for $\mathcal{C}_{f\bar{f}\rightarrow \chi\chi}$ is based on a non-relativistic expansion of the inverse reaction $\chi\chi \rightarrow f\bar{f}$ as explained in the text. Retaining the $s$-wave in the expansion of $\left.(\sigma v)_{\chi\chi \rightarrow f\bar{f}}\right|_{s=4m_\chi^2(1 + v^2/4)}$, we obtain the expression for each SM fermion but SM neutrinos
\begin{equation}
    \ev{\sigma v}_{\chi\chi \rightarrow f\bar{f}} \approx  \frac{4 N_c \, m_\chi^2 \, m_f^2 \, (m_\chi^2 - m_f^2) \, \sqrt{1 - \frac{m_f^2}{m_\chi^2}} \, (m_{h_1}^2 - m_{h_2}^2)^2 \, \sin^2\theta}{ \pi \, v_h^2 \, v_s^2\left(-4 m_\chi^2 + m_{h_1}^2 \right)^2 \left(-4 m_\chi^2 + m_{h_2}^2 \right)^2}\,,
\end{equation}
where in the case of quarks $N_c = 3$ and 1 otherwise.

\section{Thermalization considerations} \label{app:thermalization}
In this appendix, we show that the pNGB DM $\chi$ produced by freeze-in at low $\Trh$ never enters in either kinetic or chemical equilibrium. We illustrate the thermalization conditions by considering the dominant contributions in the $2\rightarrow 2$ scattering process, involving only $b$-quarks and SM Higgs decays. At $T \lesssim \Trh \simeq 5$~GeV, $2\rightarrow 2$ reaction rates involving heavy SM states ($t$, $W^\pm$, $Z$, $h$) are Boltzmann-suppressed and can be neglected, whereas lighter quarks and leptons have smaller rates due to their Yukawa suppression. 

We take two representative benchmark points that fulfill the correct relic abundance: $m_{h_2} = 10$~GeV, $\sin\theta = 10^{-4}$, $v_s = 100$~GeV, $m_\chi = 30$~GeV and $m_\chi = 64$~GeV. This choice is inspired by the target region of FCC shown in Fig.~\ref{plot_ctau2} sensitive to a long-lived $h_2$ while keeping $m_{h_2} < 2\, m_\chi$.

\subsection{Kinetic equilibrium}
Kinetic equilibrium requires that elastic scatterings between dark and SM particles occur faster than the Hubble expansion rate, $\Gamma_{\rm kin} \gtrsim H$, ensuring that the momenta follow a thermal distribution. However, in our setup, this condition is never satisfied. We illustrate this by comparing the interaction rates with the expansion rate rates. The dominant elastic scattering processes in these cases are $b(\bar{b})+\chi \rightarrow b(\bar{b})+\chi$.

\begin{figure}[t!]
    \centering
    \includegraphics[width=0.48\textwidth]{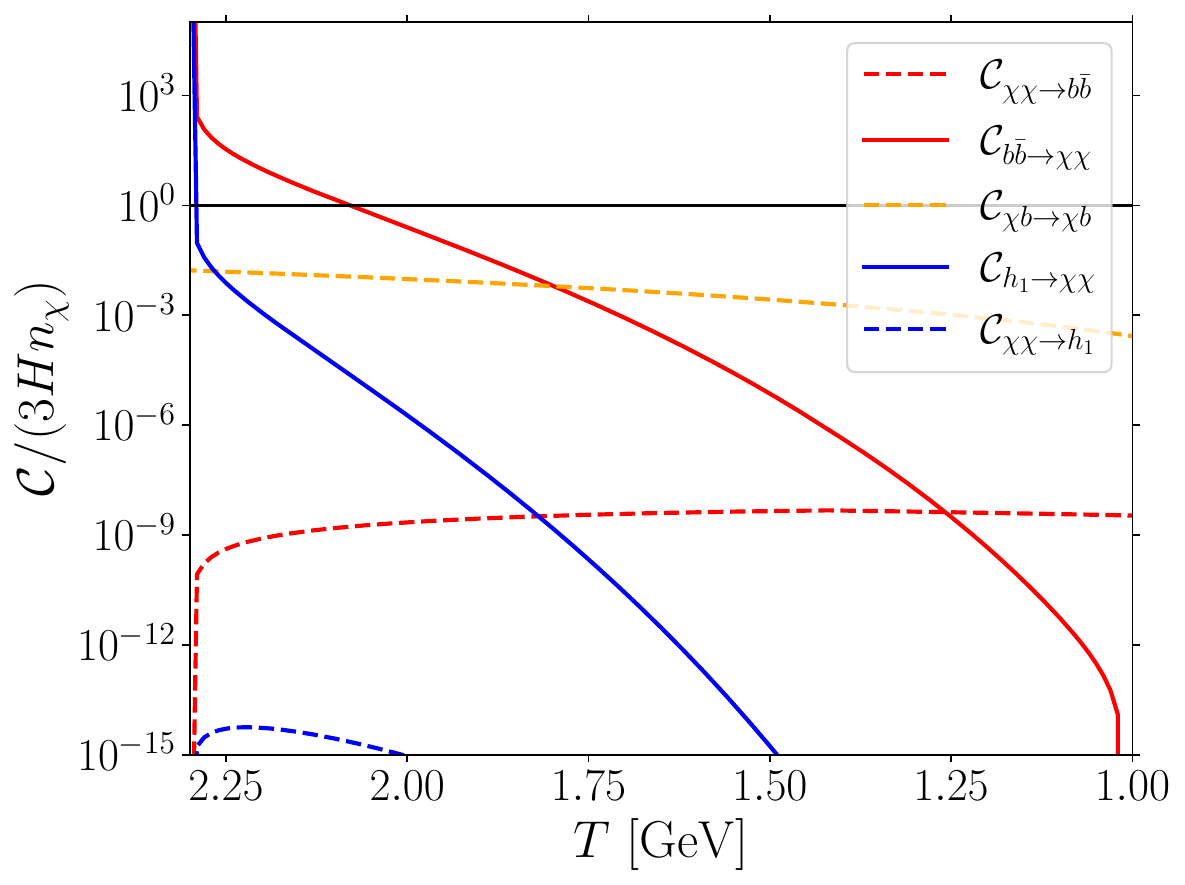} 
    \includegraphics[width=0.48\textwidth]{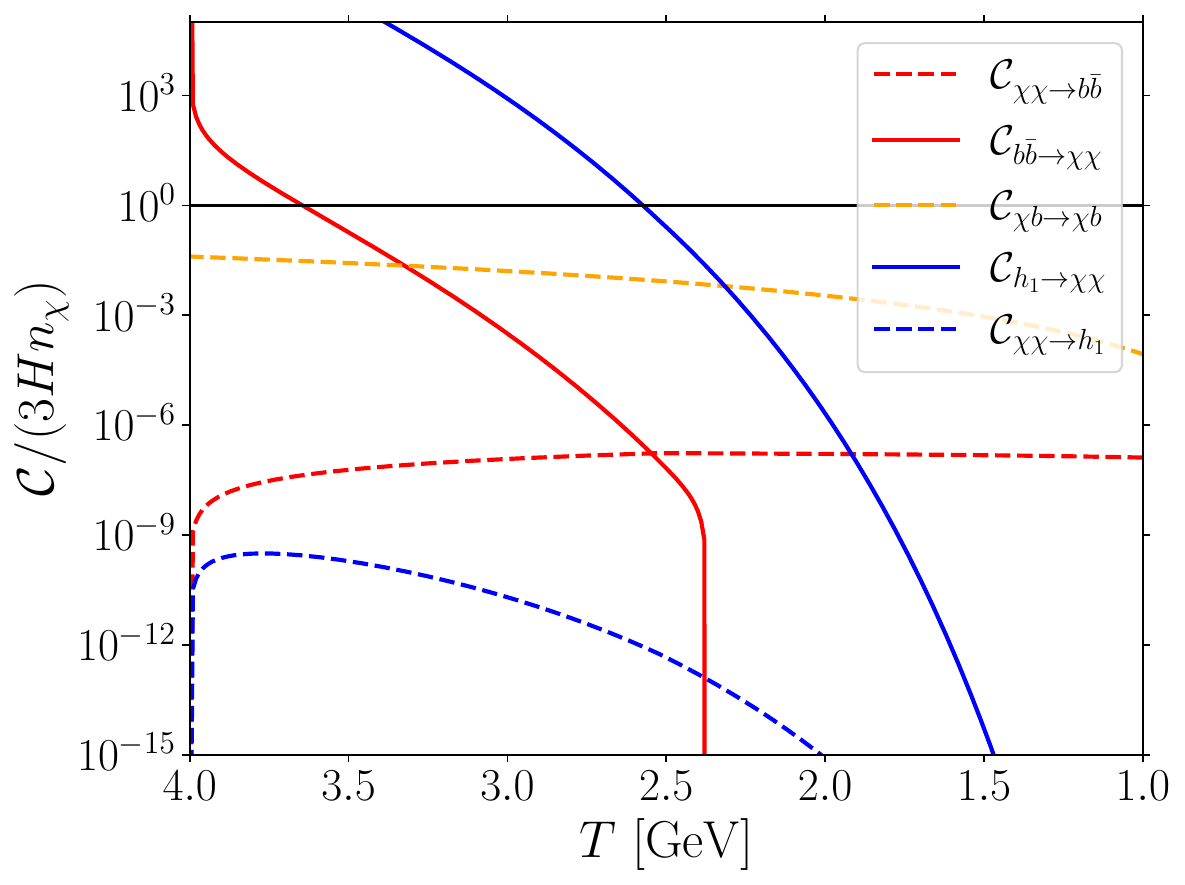} 
    \caption{Rate comparison as a function of the temperature. We have chosen $m_{h_2} = 10$~GeV, $\sin\theta = 10^{-4}$, $v_s = 100$~GeV, $m_\chi = 30$~GeV and $\Trh = 2.3$~GeV (left panel) and $m_\chi = 64$~GeV and $\Trh = 4$~GeV (right panel) (both benchmarks reproduce the whole observed DM relic abundance). The description of the different lines is given in the text.}
    \label{plot_rates}
\end{figure}

For kinetic equilibrium, we make use of the rate per unit covolume for the elastic process $\Gamma_{\rm kin} = n_{\chi}n_{b}^\text{eq} \ev{\sigma v}$, with the equilibrium number density given in Eq.~\eqref{nie}. In Fig.~\ref{plot_rates} we show with a dashed orange line the obtained $\Gamma_{\rm kin}$, which is always below $3\, H\, n_\chi$ (in both plots), indicating that the dark sector never reaches kinetic equilibrium with the SM bath. Consequently, the phase-space distributions of the two sectors evolve independently. A more refined analysis would have considered $\ev{\sigma_T v}$, with $\sigma_T$, the transfer cross section, in the calculation of the rate $\Gamma_{\rm kin}$ (times fractional momentum transfer per collision)~\cite{Profumo:2025uvx}, but the conclusion remains unchanged since $\sigma_T \leq \sigma$.
\subsection{Chemical equilibrium}
Chemical equilibrium between the two sectors could be reached if DM number-changing processes are faster than the Hubble expansion. The main processes are
\begin{align}
    \mathcal{C}_{h_1 \to \chi\chi} &= 2\, n_{h_1}^\text{eq}\, \Gamma_{h_1\to\chi\chi}\,\frac{K_1(m_{h_1}/T)}{K_2(m_{h_1}/T)}\,,\\
    \mathcal{C}_{\chi\chi\to h_1} &= \left(\frac{n_\chi}{n_\chi^{\rm eq}}\right)^2 \mathcal{C}_{h_1 \to \chi\chi}\,,\\
    \mathcal{C}_{b\bar{b} \rightarrow \chi\chi} &= \ev{\sigma v}_{b\bar{b} \rightarrow \chi\chi}\, \left({n_{\chi}^\text{eq}}\right)^2\,,\\
    \mathcal{C}_{\chi\chi\to b\bar{b}} &= \ev{\sigma v}_{\chi\chi\to b\bar{b}}\, n_{\chi}^2\,.
\end{align}

In Fig.~\ref{plot_rates} we show the ratio of the rates of the relevant processes for chemical equilibrium as a function of the temperature for the two representative benchmark points. In the left panel, the DM production is dominated by scattering (red solid) for a small fraction of temperatures until it drops below Hubble, while the inverse production is always below Hubble. In contrast, in the right panel, the Higgs decays are the main source of DM production (solid blue), whereas its inverse decay rate is always negligible. The chemical equilibrium between the two sectors is never reached and therefore DM is guaranteed to be a FIMP. We have explicitly checked that this conclusion is still valid for the parameter space relevant to the analysis.

\bibliographystyle{JHEP}
\bibliography{bibliography}

\end{document}